\documentclass[aps,twocolumn,groupedaddress,floatfix,amsmath,amssymb,pra,longbibliography]{revtex4-1}
\usepackage{amsmath}
\usepackage{amssymb}
\usepackage{graphicx}
\usepackage{textcomp}
\usepackage{bm} 
\usepackage{braket} 
\usepackage{longtable}
\usepackage[usenames,dvipsnames]{color}

\begin{document}

%\linenumbers
\title{Exploiting long-range disorder in slow-light  photonic crystal waveguides}

\author{J. P. Vasco}
\affiliation{Department of Physics, Queen's University, Kingston, Ontario, Canada, K7L 3N6}
\affiliation{Institute of Theoretical Physics, \'Ecole Polytechnique F\'ed\'erale de Lausanne EPFL, CH-1015 Lausanne, Switzerland}
\author{S. Hughes}
\affiliation{Department of Physics, Queen's University, Kingston, Ontario, Canada, K7L 3N6}

%\pacs{xxxx, xxxx, xxxx}                                         
    %\date{\today}
    
\begin{abstract} 
The interplay between order and disorder in photonic lattices opens up a wide range of novel optical scattering mechanisms, resonances, and applications that can be obscured by typical ordered design approaches to photonics. Striking examples include Anderson localization, random lasers, and visible light scattering in biophotonic structures such as butterfly wings. In this work, we present a profound example of light localization in photonic crystal waveguides by introducing  long-range correlated disorder. Using a rigorous three-dimensional Bloch mode expansion technique, we demonstrate how inter-hole correlations have a negative contribution to the total out-of-plane radiative losses, leading to a pronounced enhancement of the quality factor, $Q$, and $Q/V$ cavity figures of merit in the long-range correlation regime. Subsequently, the intensity fluctuations of the system are shown to globally increase with the correlation length, highlighting the non-trivial role of long-range disorder on the underlying scattering mechanisms. We also explore the possibility of creating ultra-high quality cavity modes via inter-hole correlations, which have various functionalities in chip-based nonlinear optics and waveguide cavity-quantum electrodynamics.
\end{abstract} 

\maketitle

\section{Introduction}
The rich physics behind phase transitions in solid state physics, such as through the metal-insulator transition \cite{tokuraRMP},  has been widely studied in electronic systems where disorder plays an important role.  The possibility of controlling phase transitions by including spatial correlations in the perturbed periodic potential has motivated systematic studies on the effects of long-range correlations to either localize or delocalize the electronic wave function throughout the system \cite{moura,carpena}. On the other hand, the interplay between order and disorder in photonics systems \cite{wiersma,disbook,wiersma2}, specifically, dielectric photonic crystal slab (PCS) waveguides, has been mainly focused during the last two decades on optimizing the performance of photonic devices where effects of disorder are usually assumed to be  detrimental \cite{povinelli,gerace1,hughes1,ramunno}, which has been shown both theoretically and experimentally. 

Despite this, there is a growing interest in employing high-quality disordered-induced localized modes on PCS waveguides, e.g., into the Anderson localization regime, to study important phenomena in cavity-QED (quantum electrodynamics) \cite{sapienza1}, collimation \cite{hsieh,bravo}, random lasing \cite{liu} and optical sensing \cite{sapienza2}, where  {extrinsic disorder} is intentionally introduced to enhance  localization, in addition to a small quantity of intrinsic disorder coming from unavoidable imperfections during the fabrication process. Furthermore, disorder has also been shown to be highly beneficial for improving a wide range of complex photonic systems with promising results \cite{bertolotti,florescu,mosk,riccardosap,sunkyu,riccardosap2,xiao}. Nevertheless, only the effects of short-range (intra-hole) correlations have been investigated in PCS devices with non-trivial results \cite{minkov1,nishan}, and the effects of long-range correlations have only been addressed in simple one-dimensional systems, including dielectric PCs \cite{liew}, metallic PCS \cite{giessen} and microwave waveguides \cite{krokhin}, whose results suggest a rich underlying physics which has hitherto remained unexplored in higher dimensions. Due to the different motivations to exploit disorder effects in photonics and electronics, a systematic study on the effects of long-range correlations on disordered PCS is now highly desired. From a theoretical perspective, this is partly because there are no well developed models to solve such a problem, as it would typically require a massive amount of computational resources, even for a single disorder configuration.

In this work, we close this gap, and report  results on the effects of inter-hole correlations on disordered-induced localized modes in disordered PCS waveguides. Specifically, we use an intuitive and fully three-dimensional Bloch mode expansion (BME) model~\cite{savona1,savona2,vasco1}, which has recently explained the observation of Anderson localization using visible light on a semiconductor chip~\cite{Crane2017}, and described how Anderson localization modes form for local disorder~\cite{vasco1}, without any fitting parameters.  Here we now significantly extend these ideas, and we show that long-range disorder is much richer and more useful in the context of light localization. Surprisingly, we find that long-range correlations induce a negative contribution to the total radiation losses of the cavity-like modes localized along the disordered waveguide, leading to a pronounced enhancements of the quality factor, $Q$, and $Q/V$ (with $V$ the effective mode volume) figures of merit commonly studied in PC cavity physics and waveguide QED. The corresponding density of states (DOS) and variance of the normalized intensity (metric commonly adopted to quantify the underlying statistical properties in disordered photonic systems) are estimated throughout a Green function formalism and found to increase (in overall) with the correlation length, thus evidencing the non-trivial role of long-range disorder on light propagation. We finally explore the possibility of creating ultra-high quality modes by taking the long-range correlated disordered dielectric profile as an optimal design for a long-length $LN$ cavity. 

Our results demonstrate
%, for first time to our knowledge, 
the rich physics arising from long-range spatial correlations in PCS systems and offer potential alternatives for using engineered disorder to create ultra-high $Q$ modes over a broad band frequency range. Moreover, they highlight a route to more easily accessing waveguide QED with quantum emitters.

\begin{figure}[t]
  \begin{center}
    \includegraphics[width=0.45\textwidth]{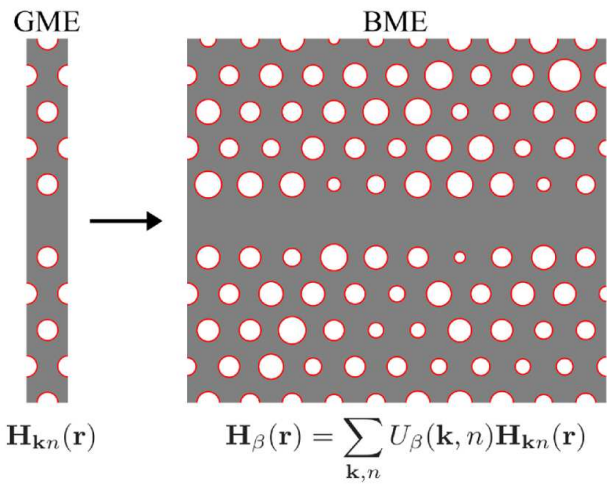}
  \end{center}
  \caption{Schematic representation of the planar system (top down view) studied in this work. The fully-vectorial Bloch modes $\mathbf{H}_{\mathbf{k}n}(\mathbf{r})$, computed with GME in a $W1$ waveguide unit cell, are used to expand the disordered modes $\mathbf{H}_{\beta}(\mathbf{r})$ with expansion coefficients $U_\beta(\mathbf{k},n)$ into the BME approximation. Radii fluctuations of different holes are assumed to be exponentially correlated with correlation length $l_c$.\label{figscheme}}
\end{figure}

\section{Theory and Computational Modelling Approach to Disordered Photonic Crystals}
From a computational perspective, modelling three-dimensional cavity modes
through long-range disorder poses a significant challenge, since even one instance requires massive amounts of memory and run time, if using a standard brute-force Maxwell
solver. Even if feasible for a few instances (which is usually not the case for a sufficient number of unit cells), full numerical simulations would offer little to no physical insight into trends and new features with correlated disorder. On the other hand, perturbative approaches would also fail~\cite{hughes1,povinelli}, as one needs a solution to the  full nonperturbative scattering problem.

To address this challenge, we investigate the disordered PCS waveguides using an efficient Bloch mode expansion method (BME) \cite{savona1}, in particular, we focus on the $W$1 system  (i.e., one row of holes removed in a triangular lattice) whose physical properties are well known. The BME approach uses the magnetic field of the disordered mode $\mathbf{H}_{\beta}(\mathbf{r})$, which is expanded in the Bloch mode basis of the non-disordered waveguide, i.e., $\mathbf{H}_{\mathbf{k}n}(\mathbf{r})$, where $\mathbf{k}$ and $n$ are the wave vector and band index, respectively, of the Bloch mode inside the projected Brillouin zone of the $W$1 system. 
Figure~\ref{figscheme} shows a schematic representation of our system. The corresponding expansion coefficients, $U_\beta(\mathbf{k},n)$, are then determined by solving Maxwell's equations, which are conveniently written as a simple linear eigenvalue problem for isotropic, non-magnetic, transparent and linear photonic systems. {Such eigenvalue problem turns to be Hermitian when working with magnetic fields instead of the electric ones, simplifying the numerical effort to solve the system}. To compute the states $\mathbf{H}_{\mathbf{k}n}(\mathbf{r})$, we apply the guided mode expansion method (GME) in which the PCS modes are expanded in the guided mode basis of the effective homogeneous slab \cite{andreani}, which is found to be very accurate for high index contrast slabs. 

For the material system,
we consider Si material parameters, with refractive index $n=3.46$, hole radii $R=0.25$ and slab thickness $d=0.55a$, where $R$ and $d$ are written in terms of the lattice parameter $a$ of the underlying hexagonal lattice of holes. The modes $\mathbf{H}_{\mathbf{k}n}(\mathbf{r})$ are accurately computed in a supercell of dimensions $a\times5\sqrt{3}a$ with one guided TE mode and a momentum cutoff $a|\mathbf{G}|_{\rm max}=19$, {where $\mathbf{G}$ is the lattice vector in the reciprocal space. This choice determines} a basis of 243 plane waves or, equivalently, 243 reciprocal lattice vectors $\mathbf{G}$. The Bloch mode expansion is then carried out utilizing the complete set of bands (243) in a disordered supercell of dimensions $100a\times5\sqrt{3}a$ and, consequently, 100 $k$ points along the 1D projected Brillouin zone of the non-disordered PCS waveguide. The radiative decay $\gamma_\beta$ of the disordered mode is numerically estimated via the photonic {\it golden rule}~\cite{andreani,savona1}, where the total ``transition probability'' from the mode $\mathbf{H}_{\beta}(\mathbf{r})$ to a radiative mode $\mathbf{H}_{\rm rad}(\mathbf{r})$ is computed and weighted with the corresponding electromagnetic density of radiative states. The mode quality factor is then $Q_\beta=\omega_\beta/\gamma_\beta$, with $\omega_\beta$ representing the resonant frequency of the $\beta$-th mode. See Supporting Information for additional details on the BME method. Disorder is introduced in the system by considering random fluctuations of the hole radii with Gaussian probability, and fluctuations from different holes are assumed to be exponentially correlated:
\begin{equation}\label{corrf}
\braket{\Delta R(\bm{\rho}_h)\Delta R(\bm{\rho}_{h'})}=\sigma_c^2e^{-\frac{|\bm{\rho}_h-\bm{\rho}_{h'}|}{l_c}},
\end{equation}
where $\sigma_c$ is the standard deviation of the Gaussian fluctuations, $\Delta R(\bm{\rho}_h)$ is the radius fluctuation of the hole positioned at $\bm{\rho}_h=(x,y)_h$ and $l_c$ is the correlation length.

\section{Properties and statistics of the disorder-induced resonances}

\begin{figure*}[t]
  \begin{center}
    \includegraphics[width=1.0\textwidth]{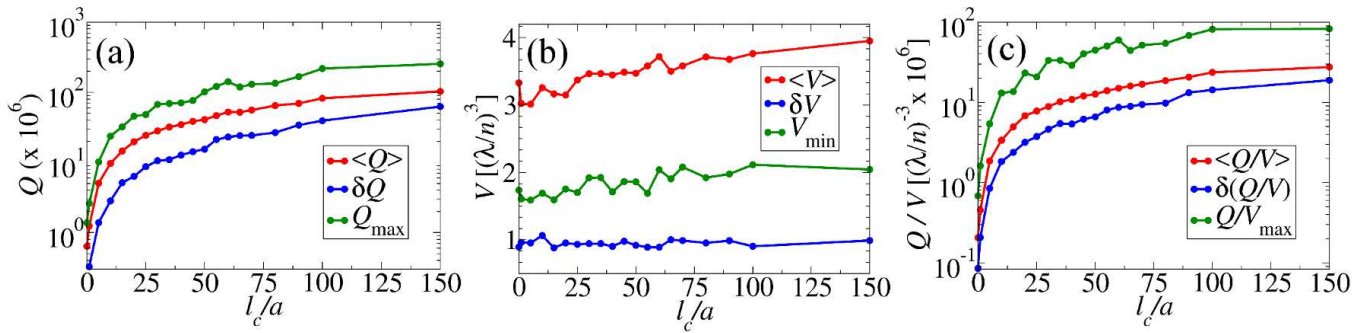}
  \end{center}
  \caption{(a) Average $\braket{Q}$ (red curve) and standard deviation $\delta Q$ (blue curve) of the cavity mode quality factors induced by disorder (in log-scale) as a function of the correlation length $l_c$. The maximum $Q$ found in the distribution is represented by the green curve. (b) Corresponding effective mode volumes from (a), with the minimum found in the distribution represented by the green curve. (c) Corresponding enhancement factor $Q/V$ from (a) and (b) with the maximum $Q/V$ found in the distribution represented by the green curve. 20 independent statistical realizations of the disordered system were considered for each $l_c$ with $\sigma_c=0.005a$ 
   \label{figmain}}
\end{figure*}

Figure~\ref{figmain}(a) shows the statistical behavior of the averaged $Q_\beta$ (in a log-scale) as a function of the correlation length for $\sigma_c=0.005a$. Supercell-induced finite size effects are avoided by simply considering only those modes whose localization length is smaller than $25a$ when computing the average, $\braket{Q}$ (red curve), and standard deviation, $\delta Q$ (blue curve). The localization length is estimated through the inverse participation number of the eigenmodes, which accurately describes the effective length cavity-like states where the mode intensity is non-negligible \cite{vasco1} (see Supporting Information). Furthermore, 20 independent statistical realizations of the disordered system are considered to compute the first and second moments of the $Q$ distributions, which is sufficient to capture the statistics of the radiation losses in the waveguide or long $LN$ cavity geometry \cite{vasco2}. The mean quality factor increases rapidly for increasing $l_c$ and starts to saturate around $l_c=100a$, yielding a $\braket{Q}$ improvement of two orders of magnitude with respect to the non-correlated case ($l_c\rightarrow 0$); the same behavior is also seen in the standard deviation $\delta Q$ and $Q_{\rm max}$ (green curve), where the latter is the largest quality factor found in the statistical ensemble (right-tail of the quality factor distribution). {Such $\braket{Q}$ improvement is also seen for larger disorder magnitudes $\sigma_c$, although smaller enhancement factors are obtained (see details in the Supporting  Information document).} The corresponding effective mode volumes, computed by means of the well known expression $V_\beta\approx 1/\mbox{Max}[\epsilon(\mathbf{r})|\mathbf{E}_{\beta}(\mathbf{r})|^2]$, are shown in Fig.~\ref{figmain}(b);
here we are treating our modes as approximate normal modes for the purpose
of normalization, which is a very good approximation for
high $Q$ modes, else one could also compute the
generalized effective mode volumes using quasinormal mode
theory~\cite{kristensen_generalized_2012}.
A slow increasing of $\braket{V}$ is seen as $l_c$ increases, leading to a maximum of $20\%$ change with respect to the non-correlated case $l_c\rightarrow0$. The minimum volume in the $V$ distribution, $V_{\rm min}$ (green curve), follows the same trend of $\braket{V}$, while the standard deviation $\delta V$ remains approximately constant around $(\lambda/n)^3$ as $l_c$ changes. The very small variations of $\braket{V}$ with respect to $\braket{Q}$ allows us to conclude, through Fig.~\ref{figmain}(c), that the enhancement factor $Q/V$ is mainly determined by the $Q$ distributions; thus, somewhat surprisingly, we conclude that {\it the cavity figure of merit $Q/V$ can be increased up to two orders of magnitude by introducing long-range correlations in the disordered waveguide.} While specific
cavity designs can also achieve very large 
$Q/V$, our emphasis here is on the rich
physics of long range disorder to create many such random resonances.

To help explain these unusual results of Fig.~\ref{figmain},  we compute the average of the radiative decay ($\gamma$) from the photonic golden rule \cite{andreani,savona1}. We first assume that the disordered modes are dominated by the Bloch mode in the band-edge of the non-disordered waveguide $\mathbf{H}_{\rm wg}(\mathbf{r})$, which is a good approximation in the weak-disorder regime. This assumption allow us to estimate the average $\gamma$ from the fluctuations of the dielectric profile only:
\begin{equation}\label{fermi1}
\braket{\gamma}=\frac{\pi}{\omega}\sum_{\rm rad}\mbox{\hspace{0.05cm}}\iint\displaylimits_{\rm S. cell}\rho_{\rm rad}(z)\braket{\eta(\mathbf{r})\eta^*(\mathbf{r}')}\Xi(\mathbf{r})\Xi^*(\mathbf{r}')d\mathbf{r}d\mathbf{r}',
\end{equation}
where $\omega$ is the frequency of the non-disordered waveguide mode at the band-edge, $\rho_{\rm rad}(z)$ is the radiative density of states of the effective homogeneous system \cite{andreani}, $\eta(\mathbf{r})=1/\epsilon(\mathbf{r})$ with $\epsilon(\mathbf{r})$ representing the disordered dielectric function, and $\Xi(\mathbf{r})=[\nabla\times\mathbf{H}^*_{\rm rad}(\mathbf{r})]\cdot[\nabla\times\mathbf{H}_{\rm wg}(\mathbf{r})]$ is the projection of the approximated disordered state on the radiative one, {which quantifies the coupling between the former and the latter, resulting in a radiative decay}. This approach is formally equivalent to the Green function formalism, using the second-order Born approximation, to compute the out-of-plane power loss in disordered PCS waveguides \cite{hughes1}, which has been successfully applied to explain various experiments~\cite{ramunno,patterson1,Patterson2009}.

The integrals in Eq.~(\ref{fermi1}) have to be computed over all the disordered supercell (S.cell) volume with $z$ ranging from $-\infty$ to $\infty$, however, since fluctuations occur solely where disorder is present, we focus on the region within the slab ($z$ ranging from $-d/2$ to $d/2$) in which the dielectric profile $\eta(\mathbf{r})$ depends on the in-plane coordinates $\bm{\rho}=(x,y)$ only and $\rho_{\rm rad}$ does not depend on $z$. Therefore, we expand $\eta(\bm{\rho})$ in a set of two-dimensional plane waves
\begin{equation}
\label{etaexp}
\eta(\bm{\rho})=\sum_{\mathbf{G}}\eta(\mathbf{G})e^{-i\mathbf{G}\cdot\bm{\rho}},
\end{equation}
where $\mathbf{G}$ is the reciprocal lattice vector of the disordered supercell. In the weak-disorder regime, i.e., $\Delta R/R \ll 1$, 
%it is easy to show that 
the first order approximation to $\eta(\mathbf{G})$ is given by
\begin{equation}\label{totalF}
 \eta(\mathbf{G})\approx\eta_0(\mathbf{G})+\alpha(\mathbf{G})\sum_{h}\Delta R(h)e^{i\mathbf{G}\cdot\bm{\rho}_h},
\end{equation}
with
\begin{align}
\eta_0(\mathbf{G})&=\frac{1}{\epsilon_s}\delta_{\mathbf{G},0}+\frac{2\pi R^2 (\epsilon_s-\epsilon_R)}{\epsilon_s\epsilon_RA_{\rm{S. cell}}}\frac{J_1(|\mathbf{G}|R)}{|\mathbf{G}|}\sum_{h}e^{i\mathbf{G}\cdot\bm{\rho}_h},\nonumber\\
\alpha(\mathbf{G})&=\frac{2\pi R (\epsilon_s-\epsilon_R)}{\epsilon_s\epsilon_RA_{\rm{S. cell}}}J_0(|\mathbf{G}|R).
\end{align}
Here, $\epsilon_s$ and $\epsilon_R$ are the dielectric constants of the slab and holes, respectively, $A_{\rm{S. cell}}$ is the area of the supercell and $J_m$ represents the Bessel function of $m$ order. In Eq.~(\ref{totalF}), the first term corresponds to the Fourier coefficient of the non-disorder profile, while the second term is the Fourier coefficient of the random fluctuations. By using Eqs.~(\ref{corrf}), (\ref{etaexp}) and (\ref{totalF}) in Eq.~(\ref{fermi1}), and considering that $\braket{\Delta R}=0$ and $\braket{\Delta R^2}=\sigma_c^2$,  we find that the average of the radiation losses can be written as
\begin{equation}\label{gamma-cont}
\braket{\gamma}=\braket{\gamma}_{\rm non{\text-}corr}+\braket{\gamma}_{\rm corr},
\end{equation}
with 
\begin{align}
 \braket{\gamma}_{\rm non{\text-}corr} =& \frac{\pi}{\omega}\sum_{\mathbf{G}\mathbf{G}'}[\eta_0(\mathbf{G})\eta_0^{*}(\mathbf{G'})+\alpha(\mathbf{G})\alpha(\mathbf{G}')\sigma_c^2 \nonumber \\
  &\times\sum_{h}e^{i(\mathbf{G}-\mathbf{G}')\cdot\bm{\rho}_h}]I_{\rm rad}(\mathbf{G},\mathbf{G}'),\label{gamma-nocorr}
\\
 %\nonumber\\
 \braket{\gamma}_{\rm corr}=&\frac{\pi\sigma_c^2}{\omega}\sum_{\mathbf{G}\mathbf{G}'}[\alpha(\mathbf{G})\alpha(\mathbf{G}')\sum_{h\neq h'}e^{-\frac{|\bm{\rho}_h-\bm{\rho}_{h'}|}{l_c}} \nonumber \\
 &\times e^{i\mathbf{G}\cdot\bm{\rho}_h}e^{-i\mathbf{G}'\cdot\bm{\rho}_{h'}}]I_{\rm rad}(\mathbf{G},\mathbf{G}'),\label{gamma-corr}
\\
%\nonumber\\
 I_{\rm rad}(\mathbf{G},\mathbf{G}')=&\sum_{\rm rad}\rho_{\rm rad}\int\int e^{-i\mathbf{G}\cdot\bm{\rho}}e^{i\mathbf{G}'\cdot\bm{\rho}'} \nonumber \\
 &\times\Xi(\mathbf{r})\Xi^*(\mathbf{r}')d\mathbf{r}d\mathbf{r}',
\end{align}
where$\braket{\gamma}_{\rm non{\text-}corr}$ represents the contribution of disorder to the total averaged out-of-plane losses with no correlations, while the term $\braket{\gamma}_{\rm corr}$ represents the contributions coming from the inter-hole correlations only. 

Since $\braket{\gamma}<\braket{\gamma}_{\rm non{\text-}corr}$, or equivalently $\braket{Q}>\braket{Q}_{\rm non{\text-}corr}=\braket{Q}(l_c\rightarrow0)$, from the results of Fig.~\ref{figmain}, we conclude that \textit{inter-hole correlations add negative contributions to the total radiative losses}. This fact, conjointly with Eq.~(\ref{gamma-cont}), are two key results of this paper. Specifically, in the regime of long-range correlations, the correlation function $\exp(-|\bm{\rho}_h-\bm{\rho}_{h'}|/l_c)$ tends to its maximum value 1, thus maximizing the negative contributions to $\braket{\gamma}$ and correspondingly maximizing the average quality factor of the disorder-induced cavities modes. The saturation of $\braket{Q}$ around $100a$ in Fig.~\ref{figmain}(a) is then understood as follows: for correlations lengths larger than the largest inter-hole separation (which is around the waveguide length $L$) the correlation function starts to saturate to its upper limit. {The dependence of our results on the specific shape of the correlation function for intermediate values of $l_c$, i.e., $l_c\lesssim L$, is not trivial and will be discussed in future work}. {It is important to say that our results in Fig.~\ref{figmain}(a) for $l_c=0$, i.e., $\braket{Q}_{l_c=0}\approx(6.4\pm1.5)\times10^5$ are in good agreement with previous experimental measurements of disorder-induced cavity modes in silicon $W1$ waveguides. Specifically, Topolancik et al.~\cite{topolancik2} have previously reported quality factors of $3\times10^4$ (largest $Q$ in the ensemble), for an intrinsic disorder magnitude of around $0.015a$. This correspond to $3\sqrt{3}$ times our $\sigma$ (the $\sqrt{3}$ comes from disorder contributions on $x$, $y$ and $r$ in the experimental sample \cite{vasco1}), which results in a factor of $1/27$ on $\braket{Q}$ (see Ref.~\citenum{momchilopex}), then yielding $\sim 2.4\times10^4$. Assuming the same scaling factor for $\delta Q$ in Fig.~\ref{figmain}(a), we finally get that, under the conditions given in Ref.~\citenum{topolancik2}, our model predicts $\braket{Q}_{l_c=0}\approx (2.4\pm 0.5)\times10^4$.}

\begin{figure}[htb!]
  \begin{center}
    \includegraphics[width=0.45\textwidth]{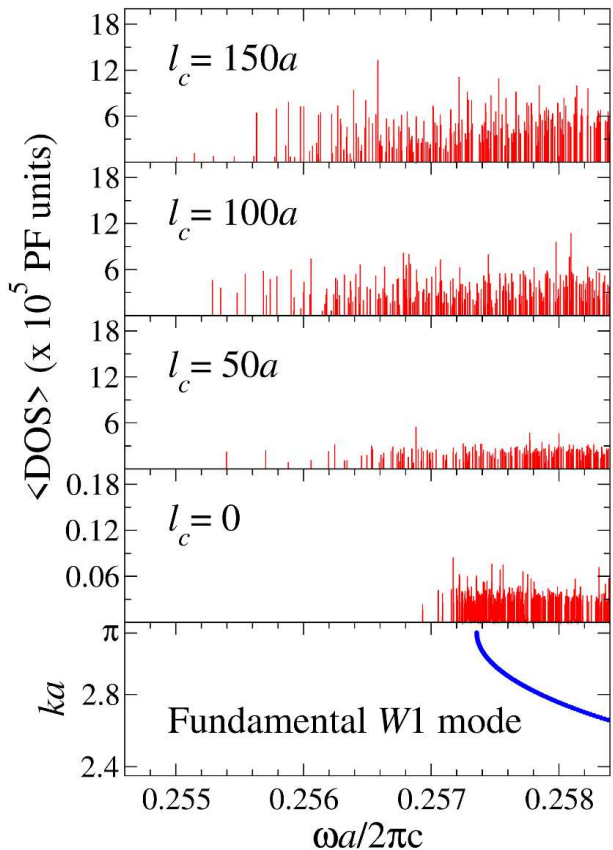}
  \end{center}
  \caption{Averaged DOS in Purcell factor units over the total number of independent disordered realizations (20 instances) as a function of frequency, for correlations lengths $l_c=0$, $l_c=50a$, $l_c=100a$ and $l_c=150a$. We also show in the bottom pannel the corresponding fundamental $W1$ mode in the same frequency range. \label{figdos}}
\end{figure}

\begin{figure}[htb!]
  \begin{center}
    \includegraphics[width=0.45\textwidth]{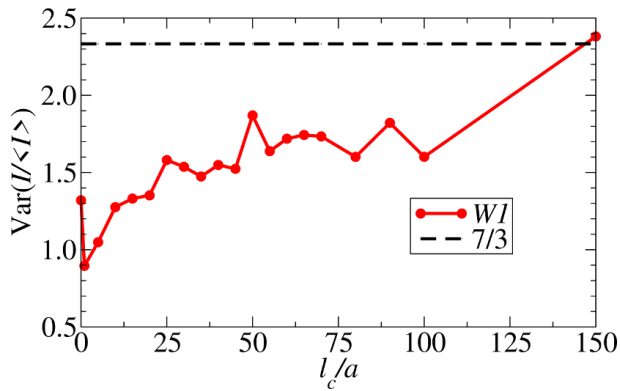}
  \end{center}
  \caption{Variance of the normalized intensity (intensity fluctuations) along the $W1$ waveguide (red line) within the same frequency range of Fig.~\ref{figdos} as a function of the correlation length. The threshold for Anderson localization is represented by the horizontal black dashed line. \label{figvarI}}
\end{figure}

To better understand the results of Fig.~\ref{figmain}, we compute the total DOS in units of the Purcell factor {for perfectly coupled emitters}, by means of the formula\cite{vasco1,vasco2}
\begin{equation}\label{DOS}
\mbox{DOS}(\omega)=\frac{6\pi c^3}{\omega\epsilon^{3/2}}\mbox{Im}\left\{\sum_{\beta}\frac{1}{\tilde{\omega}_\beta^2-\omega^2} \right\},
\end{equation}
where $\epsilon$ is the slab dielectric constant, and $\tilde{\omega}_\beta=\omega_\beta-i\gamma_\beta/2$ is the complex frequency. Equation~(\ref{DOS}) is averaged over the total number of disorder realizations and we show the corresponding $\braket{\mbox{DOS}(\omega)}$ in Fig.~\ref{figdos} for different correlation lengths. The fundamental $W$1 waveguide mode is also shown in the bottom panel of the figure, {which is within the telecom band for $a=400$~nm (see Supporting Information)}. The averaged DOS displays series of sharp spectral lines, mostly induced within the gap of the $W$1 waveguide, with intensities in the $10^6$ regime (Purcell factor units) for the longest correlation length considered. Such lines correspond to the disorder-induced cavity modes and evidence the non-trivial effects of inter-hole correlations on the multiple scattering processes into the waveguide. Particularly, an interesting trend from this figure, is the increasing number of these discrete resonances in the gap region (as well as their average height) with the correlation length, which effectively increases the operation bandwidth for applications where a large number of high-quality modes is desired over a broad frequency range, for application in random lasing and many body QED \cite{liu,hartmann,sapienza1}.

Figure~\ref{figdos} suggests an increasing of the local DOS fluctuations as the correlation length increases, since a larger number of random cavities start to appear into the system band gap. In order to verify this hypothesis we compute the intensity fluctuations which have served as a quantifier of the statistics followed by the disorder-induced resonances, and capture the transition between extended to localized states in presence of disorder \cite{Genack}. Using a photonic Green function formalism, and linking the vertically emitted radiation with the local DOS of the system \cite{Nishan2,Garcia2}, it is possible to show that the variance of the normalized intensity are given by \cite{vasco2}
\begin{equation}\label{varI}
\mbox{Var}\left[\frac{I}{\braket{I}}\right]=\mbox{Var}\left[\frac{\varrho(\mathbf{r},\omega)}{\braket{\varrho(\mathbf{r},\omega)}}\right],    
\end{equation}
where
\begin{equation}\label{lDOS}
\varrho(\mathbf{r},\omega)=\frac{6}{\pi \omega}\mbox{Im}\left\{\mbox{Tr}\left[\overleftrightarrow{\mathbf{G}}(\mathbf{r},\mathbf{r},\omega)\right]\right\},    
\end{equation}
is the local DOS \cite{novotny}, and $\overleftrightarrow{\mathbf{G}}(\mathbf{r},\mathbf{r}',\omega)$ is the transverse Green function given by \cite{Yao}
\begin{equation}\label{Gf}
\overleftrightarrow{\mathbf{G}}(\mathbf{r},\mathbf{r}',\omega)\approx \sum_\beta \frac{\omega^2\mathbf{E}_\beta^\ast(\mathbf{r}')\mathbf{E}_\beta(\mathbf{r})}{\tilde{\omega}_\beta^2-\omega^2}.
\end{equation}

Equation~(\ref{varI}) is computed for the local DOS along the waveguide direction, i.e., $\varrho(x,y=0,z=0,\omega)$, for all disordered samples and averaged over the frequency range of Fig.~\ref{figdos}. Results are shown in Fig.~\ref{figvarI}, where an overall increasing of $\mbox{Var}(I/\braket{I})$ is seen as the correlation length increases, thus confirming our hypothesis. It is worth nothing that because the very small disorder magnitude introduced in the system, a large correlation length is needed within the studied frequency window to reach the {statistical} threshold for Anderson localization, which is defined by the horizontal dashed line at 7/3 (see Ref.~\citenum{sapienza1}). {Nevertheless, sharp peaked resonances with ultra-high quality factors and localization length much smaller than the waveguide length are still found in the ensemble of disordered samples for $l_c<150a$. These disorder-induced resonances, product of multiple coherent backscattering events~\cite{Patterson2009}, are truly localized and may be employed for cavity-like physics as proposed in the following section. Yet, and more importantly,} Fig.~\ref{figvarI} showcases the profound role of the correlation length, for long-range disorder, on the transition between extended to localized photonic states, in  analogy to metal-insulator transitions in electronic systems.

\section{Using long-range correlated disorder to create ultra-high $Q$ modes}

\begin{figure}[th!]
  \begin{center}
    \includegraphics[width=0.45\textwidth]{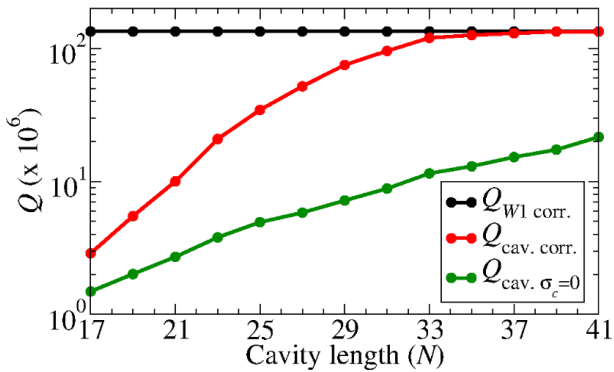}
  \end{center}
  \caption{Quality factors (log-scale) of the $LN$ cavity, $Q_{\rm cav.}^{\rm corr.}$ (red curve), designed throughout the disordered long-range correlated waveguide realization at $l_c=60a$, $Q_{W\rm 1 \text{ } corr.}$ (horizontal black curve), as a function of the cavity length. The fundamental mode $Q$ of the perfect $LN$ cavity, $Q_{\rm cav.}^{\sigma_c=0}$, is shown in green as a function of the cavity length.\label{figcav}}
\end{figure}

\begin{figure*}[th!]
  \begin{center}
    \includegraphics[width=0.99\textwidth]{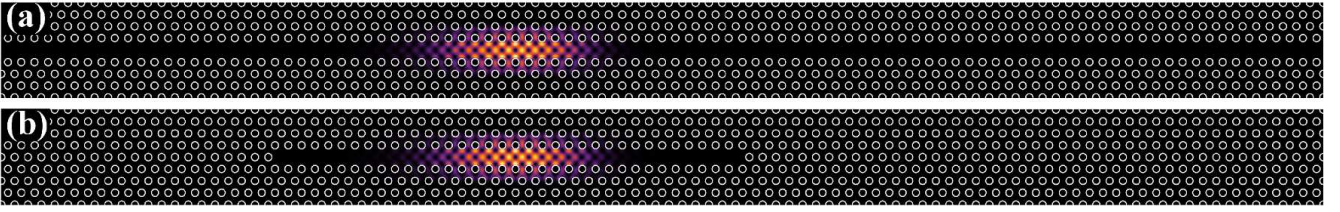}
  \end{center}
  \caption{(a) Intensity profile of the disordered-induced localized mode (within the slab center) in the $W1$ long-range correlated system chosen in Fig.~\ref{figcav}. (e) Intensity profile of the resulting $L35$ cavity.\label{figcavprof}}
\end{figure*}

The green curve in Fig.~\ref{figmain}(a) represents an efficient design to support ultra-high $Q$ cavity modes when long-range correlations are considered; in fact, it has been recently shown that disorder-induced cavity modes in waveguides may be safely bounded with mirrors with negligible effects on the resonant frequencies and quality factors \cite{vasco2}. Here, we take advantage of such phenomena, together with long-range correlations, to obtain ultra-high $Q/V$ $LN$ PCS cavities. Specifically, we select the disorder-induced cavity mode from the green curves of Figs.~\ref{figmain}(a) and \ref{figmain}(c) at $l_c=60a$ (same state), and build a $LN$ cavity (along the waveguide) whose center is as close as possible to the mode electric field peak. Figure~\ref{figcav} shows the results of this analysis. The horizontal black line, $Q_{W{\rm 1 \text{ } corr.}}$, represents the $Q$ value of the disorder-induced cavity mode in the $W1$, while the red line, $Q_{\rm cav.}^{\rm corr.}$, is the $Q$ factor of the resulting $LN$ cavity mode, as a function of the cavity length, built on the corresponding disorder realization of the long-range correlated $W1$ system, i.e., we add mirrors on the specific disordered correlated geometry. Eventually, for long cavity lengths, the disordered mode becomes insensitive to the $LN$ cavity boundary condition, and both $LN$ mode and $W1$ disorder-induced mode become equivalent \cite{vasco2}. In order to evidence the $Q$ enhancement when long-range correlations are considered, we also superimpose in Fig.~\ref{figcav} the fundamental mode quality factor of the perfect $LN$ cavity, i.e., with $\sigma_c=0$, which is denoted as $Q_{\rm cav.}^{\sigma_c=0}$. We obtain a $Q$ improvement of one order of magnitude for the longest cavities, with respect to the regular case, achieved by strategically adding inter-hole long-range correlations in the system without any optimization procedure. The intensity profiles of the disordered modes in the center of the slab ($|\mathbf{D}(\bm{\rho},z=0)|^2=\epsilon(\bm{\rho})^2|\mathbf{E}(\bm{\rho},z=0)|^2$) are shown in Figs.~\ref{figcavprof}(a) and \ref{figcavprof}(b) for the $W1$ case and resulting $L35$ cavity case, respectively. Here, one clearly sees the equivalence between the disordered modes in both waveguide and long-length cavity systems.

\begin{table*}[ht!]
 \caption{Optical properties of the fundamental $L35$ cavity with long-range correlations, and with no correlations (perfect cavity). The magnitude of non-correlated intrinsic disorder considered is $\sigma_i=0.0006a$.}
 \label{tab1}
 \begin{ruledtabular}
\begin{tabular}{cccccc}

 $Q$ & $\braket{Q}_{\sigma_i}$ & $V$  $(\lambda/n)^3$& $\braket{V}_{\sigma_i}$  $(\lambda/n)^3$ & $Q/V$  $(\lambda/n)^{-3}$ & $\braket{Q/V}_{\sigma_i}$  $(\lambda/n)^{-3}$ \\
\hline
\multicolumn{6}{c}{Fundamental $L35$ with long-range correlations} \\
%\hline
$1.26\times10^8$ & $3.52\times10^7$ & $2.25$ & $2.23$ & $5.62\times10^7$ & $1.57\times10^7$\\
\hline
\multicolumn{6}{c}{Fundamental $L35$ with no correlations} \\
%\hline
$1.3\times10^7$ & $9.98\times10^6$ & $4.62$ & $4.55$ & $2.82\times10^6$ & $2.22\times10^6$\\
\end{tabular}
\end{ruledtabular}
\end{table*}

To better connect to practical devices and real life experiments to date, we next investigate the additional effects of (unavoidable) fabrication imperfections on the correlated $L35$ cavity of Fig.~\ref{figcav} (red line). Fabrication-induced, or intrinsic disorder, is considered here by adding non-correlated Gaussian fluctuations in the hole radii of the long-range correlated design, in particular, we adopt the state-of-the-art disorder magnitude, relevant to Si fabrication techniques, which is reported to be $\sigma_i=0.0006a$ \cite{noda}, i.e., around 10 times smaller than $\sigma_c$ {(notice that the actual radial fluctuation in the presence of correlations depends on the correlation length, see Supporting Information for details)}. As in the case of correlated disorder, we consider 20 statistical realizations to compute the statistical distributions of $Q$ and $V$ induced by intrinsic imperfections. Results are summarized in Table \ref{tab1} where the averaged value over the intrinsic disorder realizations is represented by $\braket{}_{\sigma_i}$. We obtain $\braket{Q}_{\sigma_i}$ and $\braket{Q/V}_{\sigma_i}$ values which are up to 3 and 7 times larger, respectively, when the long-range correlated design is considered as the optimum one. Furthermore, these averaged values in the correlated systems are even larger than the ones obtained in the perfect case (no disorder), evidencing again, the potential benefits of engineered disorder to improve cavity figures of merit as $Q$ and $Q/V$, and demonstrating the very non-trivial role of inter hole long-range correlations.

\section{Conclusions}

We have investigated the effects of inter-hole (long range) correlations on the disordered-induced localized modes emerging in disordered PCS waveguides, using a fully three-dimensional Bloch mode expansion technique. We have demonstrated that inter-hole correlations induce negative contributions to the total averaged out-of-plane losses of these cavity-like states, thus enhancing the corresponding averaged quality factor up to two orders of magnitude with respect to the one obtained with non-correlated disorder.  We have also found that long-range correlations induce intensity fluctuations in the system, allowing the transition between extended states and Anderson localization with the correlation length as a control parameter. We then studied the possibility of creating ultra-high $Q$ and $Q/V$ cavities where the long-range correlated disordered design is taken as the optimum one without carrying out any optimization procedure. Our results open a wide range of new possibilities to explore the non-trivial effects of long-range correlations in PCS platforms. But more importantly, they demonstrate the deep connection to important problems in solid state physics and
photonic crystal physics, and clearly there remains a large body of work that could extend these general ideas to nanophotonic circuits.


\section{Acknowledgement}

This work was funded by the Natural Sciences and Engineering Research Council of Canada, and Queen’s University, Canada. We gratefully acknowledge Nishan Mann for useful discussions. This research was enabled in part by computational support provided by the Centre for Advanced Computing (http://cac.queensu.ca) and Compute Canada (www.computecanada.ca).

\onecolumngrid

\appendix

\section{The Bloch Mode Expansion Method}
The eigenfrequencies and eigenstates of the disordered $W1$ system are computed in the main manuscript using the Bloch mode expansion method (BME) \cite{savona1,savona2}. The key idea behind the method is to express the disordered magnetic field of the photonic crystal, $\mathbf{H}_{\beta}(\mathbf{r})$, as a linear expansion of the non-disordered Bloch modes, $\mathbf{H}_{\mathbf{k}n}(\mathbf{r})$, with expansion coefficients $U_{\beta}(\mathbf{k},n)$:
\begin{equation}\label{bmeexpansion}
\mathbf{H}_{\beta}(\mathbf{r})=\sum_{\mathbf{k},n}U_{\beta}(\mathbf{k},n)\mathbf{H}_{\mathbf{k}n}(\mathbf{r}),
\end{equation}
where $\mathbf{k}$ and $n$ denote the wave vector of the Bloch mode within the first Brillouin zone of the non-disordered system, and $n$ labels its band index. Under the conditions of linear, isotropic, non-magnetic, transparent, and non-dispersive materials, the time-independent Maxwell equations are turned into the following ordinary linear eigenvalue problem when the expansion of Eq.~(\ref{bmeexpansion}) is considered:
\begin{equation}\label{eigenBME}
\sum_{\mathbf{k},n}\left[V_{\mathbf{k}n,\mathbf{k}'n'}+\frac{\omega^2_{\mathbf{k}n}}{c^2}\delta_{\mathbf{k}\mathbf{k}',nn'}\right]U_{\beta}(\mathbf{k},n)=\frac{\omega^2_{\beta}}{c^2}U_{\beta}(\mathbf{k'},n'),
\end{equation}
with disordered matrix elements given by
\begin{equation}\label{matrixV}
V_{\mathbf{k}n,\mathbf{k}'n'}=\int_{\rm s. cell}\delta\eta(\mathbf{r})\left(\nabla\times\mathbf{H}_{\mathbf{k}n}(\mathbf{r})\right)\cdot\left(\nabla\times\mathbf{H}^{\ast}_{\mathbf{k}'n'}(\mathbf{r})
\right)d\mathbf{r}.
\end{equation}
The matrix elements $V_{\mathbf{k}n,\mathbf{k}'n'}$ are computed in a large supercell characterized by the dielectric profile $\delta\eta(\mathbf{r})$ which is defined as
\begin{equation}\label{etadef}
\delta\eta(\mathbf{r})=\eta(\mathbf{r})-\eta_0(\mathbf{r})=\frac{1}{\epsilon(\mathbf{r})}-\frac{1}{\epsilon_0(\mathbf{r})},
\end{equation}
where $\epsilon(\mathbf{r})$ and $\epsilon_0(\mathbf{r})$ are the dielectric function of the disordered and non-disordered systems, respectively. In order to efficiently compute the integrals of Eq.~(\ref{matrixV}), we adopt the GME approximation \cite{andreani} where the Bloch modes of the non-disordered structure are expanded in the eigenmodes of the effective homogeneous slab. Since these eigenmodes are linear combinations of sine and cosine functions, the matrix elements of Eq.~(\ref{matrixV}) are turned into analytical expressions under the GME approach. Finally, the radiation mode losses of the system is estimated by computing the first order coupling between the disordered and radiative modes. This procedure is formally equivalent to the well known Fermi's golden rule in quantum mechanics, and has been shown to be very accurate in low-loss photonic crystal slabs \cite{andreani}. Specifically, the out-of-plane losses is calculated by computing the ``transition probability'' from a disordered mode $|\mathbf{H}_{\beta}\rangle$ to a radiative mode $|\mathbf{H}_{\rm rad}\rangle$, above the light line of the slab, with the corresponding radiative density of states $\rho_{\rm rad}(z)$, so that
\begin{equation}\label{photonicrule}
 \gamma_\beta = \frac{\pi}{\omega_\beta}\sum_{\rm rad} \left|\langle\mathbf{H}_{\rm rad}|\hat{\Theta}|\mathbf{H}_{\beta}\rangle\right|^2\rho_{\rm rad}(z),
\end{equation}
where 
\begin{equation}
\hat{\Theta}=\nabla\times\eta(\mathbf{r})\nabla\times,
\end{equation}
is the Maxwell operator of the disordered supercell. The out-of-plane quality factor of the optical cavity mode is defined as $Q_\beta=\omega_\beta/\gamma_\beta$, and the electric field of the disordered mode is computed from $\mathbf{H}_{\beta}(\mathbf{r})$ via Maxwell's equations, i.e., 
\begin{equation}\label{efield}
\mathbf{E}_{\beta}(\mathbf{r})=\frac{ic}{\omega_\beta\epsilon(\mathbf{r})}\nabla\times\mathbf{H}_{\beta}(\mathbf{r}),
\end{equation}
which is normalized through
\begin{equation}\label{enorm}
\int_{\rm s. cell}\epsilon(\mathbf{r})|\mathbf{E}_{\beta}(\mathbf{r})|^2d\mathbf{r}=1.
\end{equation}
Since we are dealing with linear optics and low-loss dielectric systems, we adopt the usual definition of the effective mode volume $V_\beta\approx 1/\mbox{Max}[\epsilon(\mathbf{r})|\mathbf{E}_{\beta}(\mathbf{r})|^2]$, where $\mathbf{r}_0$ is taken at the position of the electric field peak. The cavity modes are convergent
in space through the BME approach, and so we do not need
to use a formal quasinormal mode theory of these
modes~\cite{kristensen_generalized_2012}.

\subsection{Guided mode expansion band structure of perfect lattice (no disorder)}

\begin{figure}[ht!]
\centering
\includegraphics[width=0.8\textwidth]{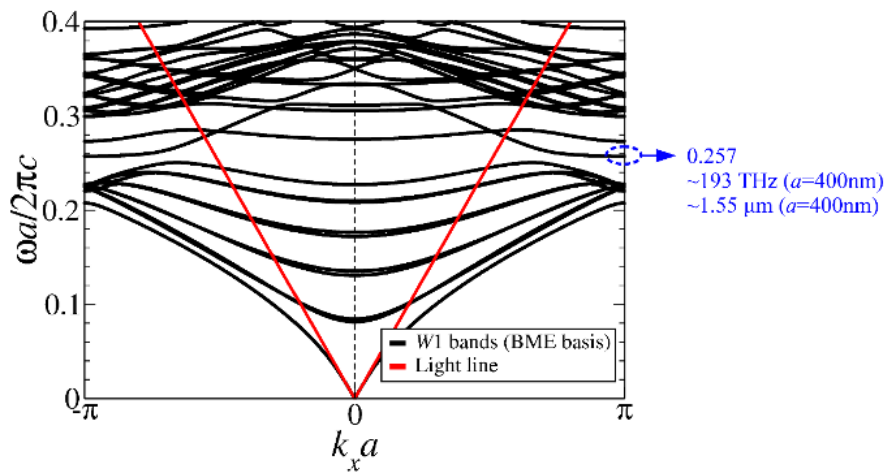}
\caption{GME projected band structure of the $W1$ waveguide. All the set of bands (black curves) are employed in the BME expansion. The band edge of the fundamental $W1$ mode is at $0.257$~$\omega a/2\pi c$, or $\nu \simeq 193$~THz / $\lambda_0 \simeq 1.55$~$\mu$m for $a=400$~nm.\label{fig1}}
\end{figure}

Figure~\ref{fig1} shows the projected band structure of the $W1$ waveguide computed with GME in the first Brillouin zone. As reported in the main manuscript, we have considered a Si refractive index of $n=3.46$, hole radii $R=0.25a$, slab thickness $d=0.55a$ and supercell dimensions of $a\times5\sqrt{3}a$. One TE guided mode is considered in the GME expansion with 243 plane waves, leading to a total set of 243 bands (partly shown by the black curves in the figure). The band edge of the fundamental $W1$ TE-like mode is found at $0.257$~$\omega a/2\pi c$, which is equivalent to $\nu \simeq 193$~THz or $\lambda_0 \simeq 1.55$~$\mu$m (telecom wvalengths) when a typical lattice parameter in Si photonic structures, $a=400$~nm, is considered. We employ the complete set of bands in the BME expansion in order to obtain accurate values for the out-of-plane losses. Further details on the convergence of the BME method can be found in Refs.~\onlinecite{momchilopex,vasco2}.

\subsection{Localization length of disorder-induced modes}

Structural disorder in PCS waveguides dramatically affects the propagation of guided modes by inducing multiple backscattering phenomena~\cite{patterson1}, which leads to the spontaneous formation of random cavities along the waveguide direction. The effective length of such cavities is determined by the effective localization length of its modes, i.e., the size of the region where the mode intensity is non-negligible, and it is a measure of the localization properties of the system in the presence of disorder. The localization length of a disordered eigenmode $\mathbf{H}_{\beta}(\mathbf{r})$ can be easily estimated in the BME formalism by means of the inverse participation number \cite{savona1}. If we define the function
\begin{equation}
\psi(x)=\int_{L_y}|\mathbf{H}_{\beta}(x,y,z=0)|dy,
\end{equation}
where $L_y$ is the size of the supercell in the $y$ direction, the inverse participation number (with length units) can be defined through
\begin{equation}\label{invpn}
L_\beta=\frac{\left[\int_L\psi(x)dx\right]^2}{\int_L\psi(x)^4dx},
\end{equation}
which accounts for the localization length of the disordered mode $\mathbf{H}_{\beta}(\mathbf{r})$ in a waveguide of length $L$.

\section{Dependence of $\bm{Q}$ on the correlated-disorder parameter $\bm{\sigma_c}$}
\begin{figure}[ht]
\centering
\includegraphics[width=0.5\textwidth]{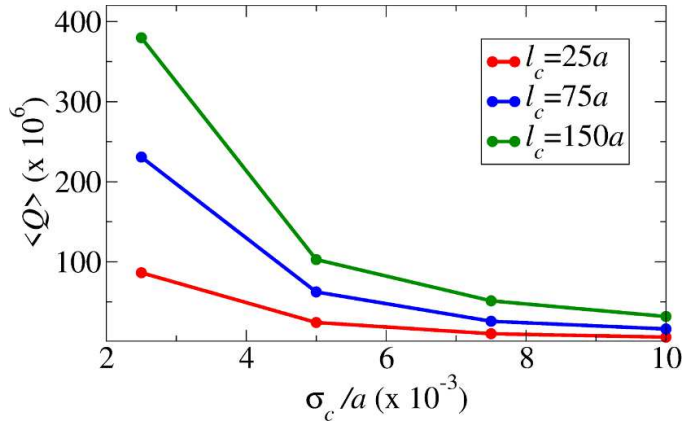}
\caption{Average $\braket{Q}$ as a function of $\sigma_c$ for different correlation lengths. The value considered in the main manuscript is $\sigma_c/a=5\times10^{-3}$. 20 independent realizations of the disordered system were considered to compute the average. \label{fig2}}
\end{figure}

In Fig.~\ref{fig2} we show the dependence of the averaged $Q$ on the correlated disorder magnitude $\sigma_c$. We have identified the same exponential decreasing of $\braket{Q}$ already seen for the Anderson localized modes in a disordered $W1$ with no spatial correlations \cite{vasco1}. As expected, the decreasing is faster for systems with very large $Q$ and huge values of $\braket{Q}$, larger than the ones reported in the main manuscript, are predicted for $\sigma_c<0.005a$. Nevertheless, $\sigma_c=0.005a$ corresponds to a good compromise between $Q$ enhancement and sample fabrication issues, as this correlated disorder magnitude is around 10 times larger than the state-of-the-art precision achieved in fabricated Si PCS structures, which is $\sigma_i\simeq0.0006a$ \cite{noda}. Notice, however, that the effective fluctuations of the holes may be smaller than the nominal $\sigma_c$ for long correlations lengths (see following section), but in the largest $l_c$ case considered these effective fluctuations are still around 5 times larger than $0.0006a$.

\section{Effective disordered system with no spatial correlations}
\begin{figure}[ht!]
\centering
\includegraphics[width=0.95\textwidth]{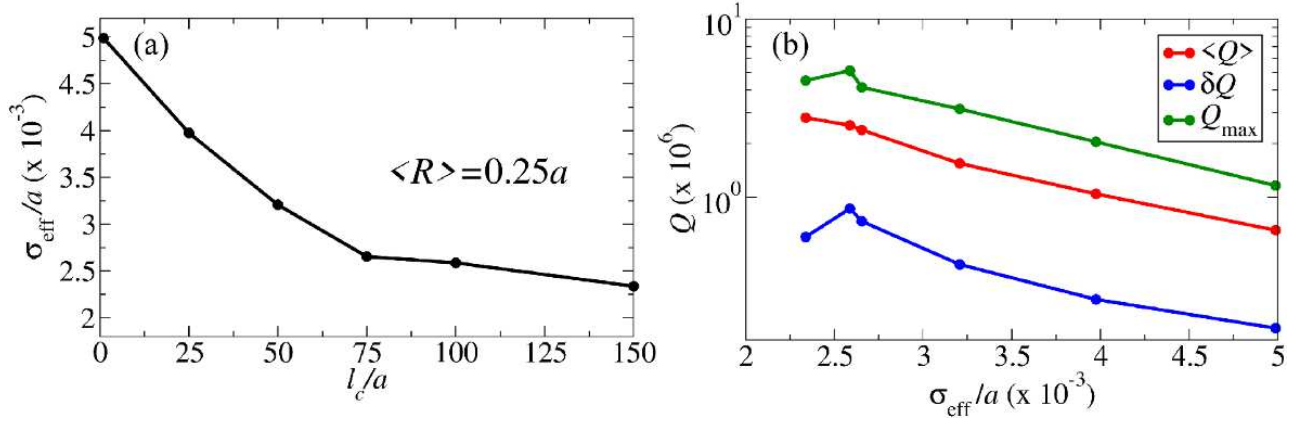}
\caption{(a) Effective radii fluctuations $\sigma_{\rm eff}$ as a function of the correlation length. The average of $R$ is found to be $0.25a$ in all cases. (b) Average $\braket{Q}$, standard deviation $\delta Q$ and maximum $Q_{\rm max}$ as a function of the effective radii fluctuations $\sigma_{\rm eff}$. 20 independent realizations of the disordered system (with no correlations) were considered to compute the average and standard deviation in panel (b) for each $\sigma_{\rm eff}$.\label{fig3}}
\end{figure}

When long correlations lengths are considered in the disordered system, the effective radii fluctuations $\sigma_{\rm eff}$ in each disordered sample is smaller than the nominal value $\sigma_c$. Such effect is clearly seen in Fig.~\ref{fig3}(a) where we plot the averaged standard deviation of $R$ over the set of disorder realizations of the system, i.e., we compute the standard deviation of $R$ for each of the 20 disordered instances and the mean of these 20 values is defined as $\sigma_{\rm eff}$. For $l_c=0$ the effective fluctuations are the same of the nominal value considered in the main manuscript, i.e., $\sigma_c=0.005a$, but for $l_c=150$ they are reduced by 50\%. In all cases the average $\braket{R}$ has the same value of the nominal hole radius $R=0.25a$. Since an effective reduction of the radii fluctuations in a disordered PCS waveguide leads to a reduction of out-of-plane losses, it might be argued that the $Q$ enhancement presented in Fig.~1(a) of the main manuscript comes from an effective reduction of structural disorder and not from the effects of long-range correlations. In order to check the robustness of our findings and discard this possibility, we investigated the behavior of $Q$ as a function of $\sigma_{\rm eff}$ with no correlations. As for the cases presented in the main manuscript, we considered 20 disorder realizations for each $\sigma_{\rm eff}$ to carry out this analysis. Results are shown in Fig.~\ref{fig3}(b). The averaged quality factor of the disorder-induced cavities modes decreases for increasing $\sigma_{\rm eff}$ as intuitively expected (increasing disorder), but the largest $\braket{Q}$ value, which corresponds to the smallest $\sigma_{\rm eff}$, is around 50 times smaller than one obtained with $\sigma_c=0.005a$ and long-range correlations [case of $l_c=150a$ in Fig.~1(a) of the main manuscript]. Therefore, results of Fig.~\ref{fig3} strongly supports the (significantly) non-trivial effects of long-range correlations in our system.

\section{Long-range correlations in the hole positions}

\begin{figure}[ht!]
\centering
\includegraphics[width=0.5\textwidth]{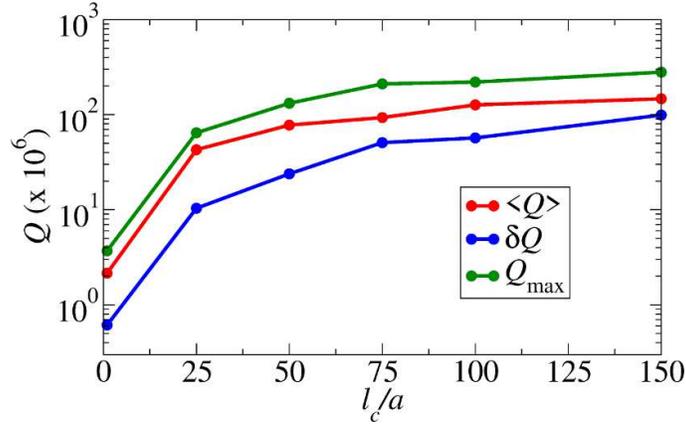}
\caption{Average $\braket{Q}$ (red curve) and standard deviation $\delta Q$ (blue curve) of the cavity mode quality factors induced by hole-position disorder (in log-scale) as a function of the correlation length $l_c$. The maximum $Q$ found in the distribution is represented by the green curve. 20 independent statistical realizations of the disordered system were considered for each $l_c$ with $\sigma_c=0.005a$.\label{fig4}}
\end{figure}

Our main conclusion in the main manuscript comes from a disorder model in which only the size of the holes is randomly fluctuated with long-range correlations. However, our findings are general and also apply for disorder in the hole positions. In such a case, we simply assume that the position random fluctuations, generated by a Gaussian probability distribution, are correlated as follows
\begin{equation}\label{corrfxy}
\braket{\Delta X_i(\bm{\rho}_h)\Delta X_i(\bm{\rho}_{h'})}=\sigma_c^2e^{-\frac{|\bm{\rho}_h-\bm{\rho}_{h'}|}{l_c}},
\end{equation}
where $X_1(\bm{\rho}_h)=x_h$ and $X_2(\bm{\rho}_h)=y_h$ are the in-plane coordinates of the hole $h$. Results are shown in Fig.~\ref{fig4} for $\braket{Q}$, $\delta Q$ and $Q_{\rm max}$ as a function of the correlation length with $\sigma_c=0.005a$. We identify exactly the same $Q$ enhancement seen for hole-size disorder in the main manuscript, strengthening the generality of our results.

\section{Robustness of the long-range correlated designs against intrinsic disorder}

\begin{figure}[ht!]
\centering
\includegraphics[width=0.5\textwidth]{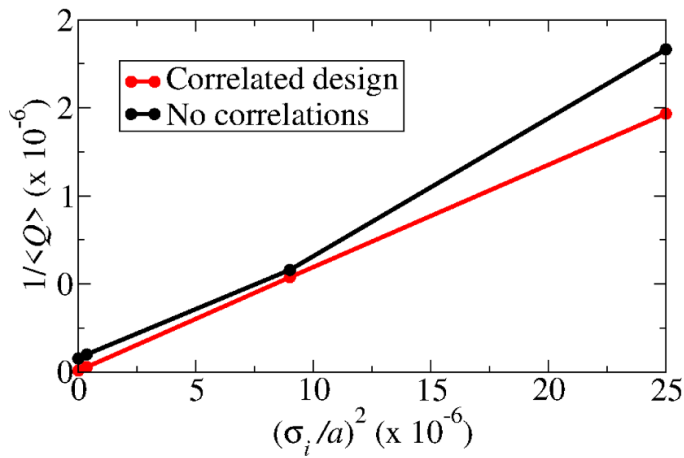}
\caption{Out-of-plane losses of the fundamental $L35$ cavity mode as a function of the squared intrinsic disorder parameter for the selected design presented in the main manuscript (red courve). The perfect case, i.e., cavity with no correlations, is also shown (back curve). We consider 20 independent statistical realizations of the disordered $L35$ cavity  for each $\sigma_i$.\label{fig5}}
\end{figure}

Figure~\ref{fig5} shows the dependence of the fundamental mode quality factor of the $L35$ design with long-range correlations (Table 1 main manuscript) on the intrinsic disorder parameter $\sigma_i$. We plot $1/\braket{Q}$ as a function of $\sigma_i^2$ in order to evince the well known linear relation between the averaged out-out-plane losses and the squared disorder parameter in PCS cavities \cite{momchilopex}. The slope of these curves are closely related with the robustness of the $Q$ cavity modes against disorder; smaller the slope more robust the system is against structural imperfections unintentionally introduced during the fabrication stage. Interestingly, the slope of the black curve is $\sim20\%$ larger than the red one, suggesting that long-range correlations may help to improve the robustness of the system against non-correlated disordered.

%\newpage

\twocolumngrid

\end{document}